\documentclass[aps,prl,twocolumn,showpacs]{revtex4-1}
\usepackage{amsfonts}
\usepackage{amsmath}
\usepackage{amsthm}
\usepackage{graphics}
\usepackage{graphicx}
\usepackage{subfigure}
\usepackage{amssymb}
\usepackage{graphics}
\usepackage{graphicx}
\usepackage{epstopdf}
\usepackage{color}
\usepackage{subfigure}
\usepackage{pgfplots,tikz}
\usepackage{hyperref}
\usepackage{soul}
\newcommand\dertt[1]{ \frac{\partial{ #1}}{\partial t} }

\DeclareMathOperator{\arccot}{arccot}

\begin{document}

\title{Irreversible dynamics of vortex reconnections in quantum fluids}

\author{Alberto Villois}
\affiliation{Department of Physics, University of Bath, Bath, BA2 7AY, UK}
\affiliation{School of Mathematics, University of East Anglia, Norwich Research Park, Norwich NR4 7TJ, United Kingdom}

\author{Davide Proment}
\affiliation{School of Mathematics, University of East Anglia, Norwich Research Park, Norwich NR4 7TJ, United Kingdom}

\author{Giorgio Krstulovic}
\affiliation{Universit\'e C\^ote d'Azur, Observatoire de la C\^ote d'Azur, CNRS, Laboratoire Lagrange, Bd de l'Observatoire, CS 34229, 06304 Nice cedex 4, France.}

%\pacs{67.25.dk, 47.37.+q, 67.25.dt, 03.75.Kk }

\begin{abstract}
We statistically study vortex reconnections in quantum fluids by evolving different realizations of vortex Hopf links using the Gross--Pitaevskii model.
Despite the time-reversibility of the model, we report a clear evidence that the dynamics of the reconnection process is time-irreversible, as reconnecting vortices tend to separate faster than they approach.
Thanks to a matching theory devised concurrently in \emph{Proment and Krstulovic (2020)} \cite{Proment2020Matching}, we quantitatively relate the origin of this asymmetry to the generation of a sound pulse after the reconnection event.
Our results have the prospect of being tested in several quantum fluid experiments and, theoretically, may shed new light on the energy transfer mechanisms in both classical and quantum turbulent fluids.  
\end{abstract}
\maketitle

%\section{Introduction}
\noindent{\it Introduction}.
Irreversibility emerges naturally in most interacting systems characterized by a huge number of degrees of freedom.
Its manifestation is associated to a time-symmetry breaking: the arrow of time appears inherently defined in the dynamics and an experienced observer is able to distinguish what are {\it before} and {\it after}.

In dissipative systems the arrow of time naturally reflects the dynamics that minimizes the energy.
Classical viscous fluids present valuable examples.
{When} no external forces are applied, an initial laminar flow decays in time until its kinetic energy is totally converted into heat. 
A less simpler example is the particle pair dispersion in turbulent flows. Although two tracers separate from each other backward and forward in time with the same Richardson scaling, their rates are different \cite{SalazarCollinsRelativeDisp}: particles separate slower forward in time than backward. 

Conservative (energy-preserving) systems are more subtle.
The arrow of time is defined only in a statistical sense by exploiting an entropy function that approaches its extremal as time progresses.
The simplest example of this kind is the free expansion experiment of a gas: even if the gas particles interact microscopically through conservative collisions, on average their macroscopic position and velocity distribution obeys the Boltzmann kinetic equation which is time-irreversible. 

\begin{figure*}
\includegraphics[width=0.99\textwidth]{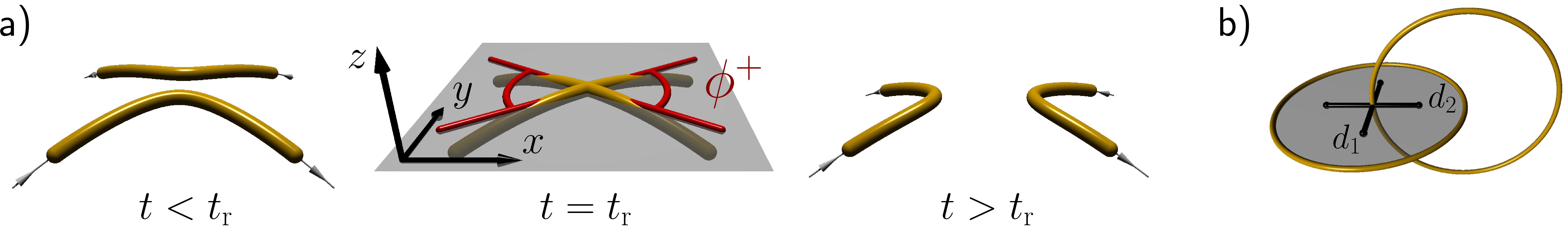} 
\caption{(Color online) 
(a) Sketch of a vortex reconnection event in quantum fluids: at the reconnection time $ t_{\rm r} $ the reconnecting filaments are locally tangent to the plane $ xOy $, here depicted in grey, and form the reconnecting angle $ \phi^+ $.
The vorticity of the filaments is depicted with grey arrows. 
(b) The Hopf link initial condition used to create the different realizations, with visual indication of the offset parameters $ (d_{1}, d_{2}) $.}
\label{fig1}
\end{figure*}

Quantum fluids are exotic types of fluids characterized by the total absence of viscosity, thus being conservative. 
Examples of such systems are superfluid liquid helium \cite{Donnelly:1991aa} and Bose--Einstein condensates (BECs) made of dilute gases of bosons \cite{Pitaevskii:2016aa}, Cooper-paired fermions \cite{Leggett:2006aa}, or massive photons \cite{Carusotto:2013aa}.
As a consequence of the wave nature of their bosonic constituents, quantum fluids have two striking properties: vortices arise as topological defects in the order parameter and their circulation takes only discrete multiples of the quantum of circulation $ \Gamma = h/m $, where $ h $ is the Planck constant and $ m $ is the boson's mass.
These defects, referred in the following as vortex filaments, present a complicate dynamics whose still misses a general solution. 
A key point in such dynamics is the occurrence of reconnection events.
A vortex reconnection is the process of interchange of two sections of different filaments, see a sketch in Fig.~\ref{fig1}(a).
It happens at small spatial and fast time scales \footnote{small length scales compared to the average filaments' length and fast time scales compared to the average filaments' length divided by their average speed}, and allows the filament topology to vary.

For the sake of simplicity, we consider in this Letter a quantum fluid described by a single scalar order parameter. 
In the limit of zero temperature, this quantum fluid accommodates only two distinct excitation families: vortex-type excitations, in the form of filaments, and compressible density/phase excitations, that is sound waves.
While the full dynamics is energy-preserving, the energy may continually flow between these two excitation families.
In this perspective, we provide a statistical analysis over many realizations of vortex reconnections, unveiling an inherent irreversible dynamics of the reconnection process.
Moreover, we show how the linear momentum and energy transfers, from vortex-type excitations to compressible density/phase excitations, is related to the geometrical parameters (macroscopic reconnection angle and concavity parameter) of the reconnecting filaments, explaining the origin of such irreversibility.

%\section{The model}
\noindent{\it Main results}.
We choose an initial configuration characterized by a Hopf link vortex filament, see Fig.\ref{fig1}(b), where (almost) all the superfluid kinetic energy is stored into the vortex-type excitations.
Similarly to vortex knots, the Hopf links naturally decay into topologically simpler configurations \cite{PhysRevE.85.036306, Proment_2014, Kleckner:2016aa} by performing a set of vortex reconnections.
{To study the Hopf link evolution, we use the Gross--Pitaevskii (GP) model,} a nonlinear partial differential equation formally derived to mimic the order parameter $\psi$ of a BEC made of dilute locally-interacting bosons, but qualitatively able to mimic a generic quantum fluid \cite{PhysRevE.93.061103}.
The GP equation, casted in terms of the healing length $\xi$ and the sound velocity $c$, reads
\begin{equation}
i\dertt{\psi} = \frac{c}{\sqrt{2}\xi}\left(-\xi^2\nabla^2\psi+\frac{m}{\rho_0}|\psi|^2\psi \right) \, ,
\label{Eq:GPhydro}
\end{equation}
where $\rho_0$ is the bulk superfluid density and $m$ the mass of a boson. 
When the GP equation is linearized about the uniform bulk value $ |\psi_0|= \sqrt{\rho_0/m} $, dispersive effects arise at scales smaller than $\xi$ and (large-scale) sound waves effectively propagate at speed $c$.
In this Letter lengths and times are expressed in units of $\xi$ and $\tau=\xi/c$, respectively. 
Thanks to the Madelung transformation
$ \psi({\bf x},t)=\sqrt{\rho({\bf x},t)/m} \exp[i \phi({\bf x},t)/(\sqrt{2}c\xi) ] $, eq.~(\ref{Eq:GPhydro}) can be interpreted as a model for an irrotational inviscid barotropic fluid of density $ \rho $ and velocity $ \mathbf{v}=\nabla\phi $.  
Vortices arise as topological defects of circulation $ \Gamma=h/m=2\sqrt{2}\pi c\xi $ and vanishing density core size order of $ \xi $ \cite{pitaevskii1961vortex}. In the previous formula, $h$ is the Planck constant.

\begin{figure}
\includegraphics[width=0.99\columnwidth]{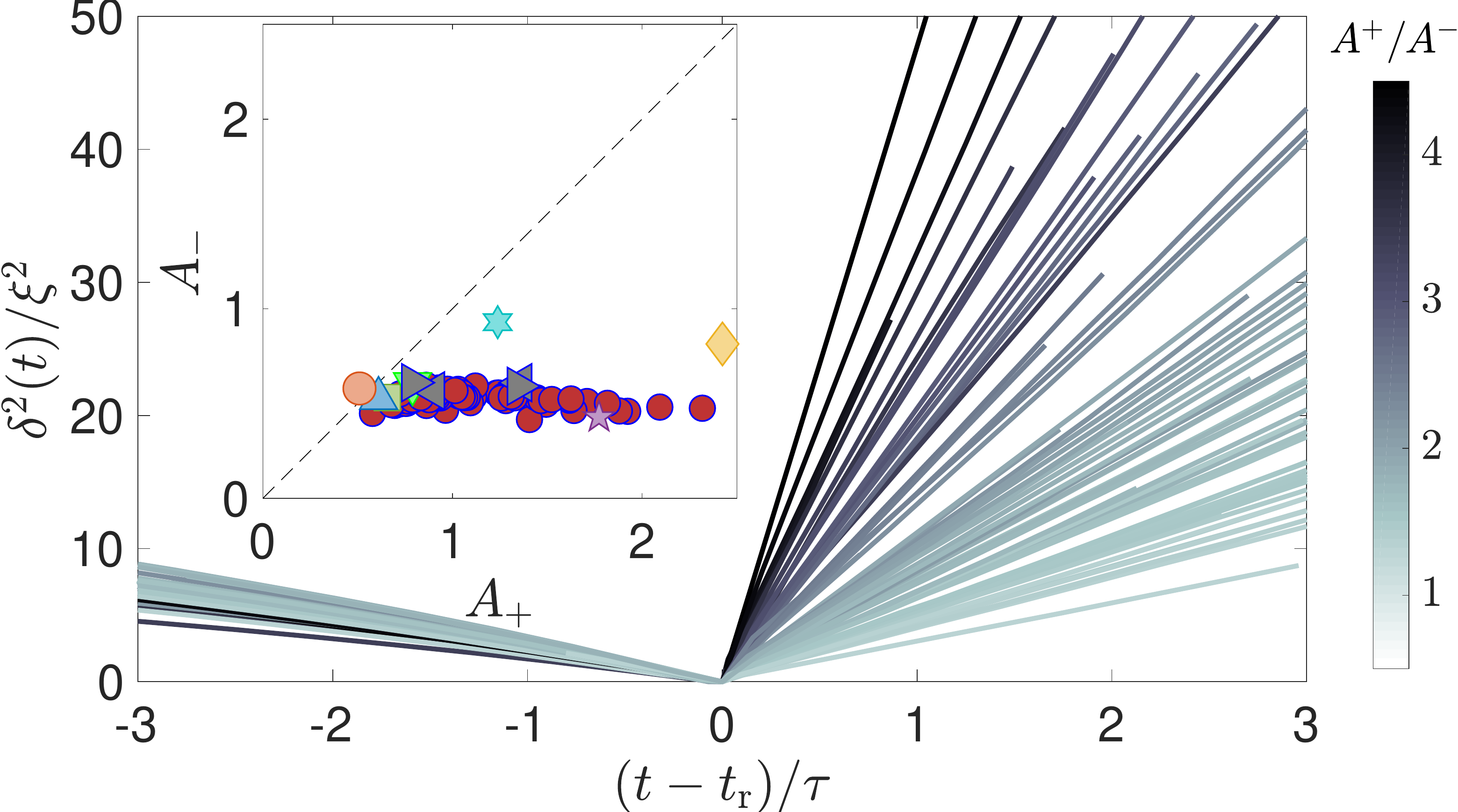}%DeltasAmApbis
\caption{(Color online) 
Squared distances versus time of the reconnecting filaments measured in all the 49 realizations.
The grey-scale color indicate the estimated value of $ A^+/A^- $ in each case.
(inset) Values of approach and separation pre-factors $A^+$ and $A^-$. Red points correspond to data of the present work. Gray left and right triangles correspond to reconnections of free and trapped vortices respectively, from Galantucci et al.\cite{galantucci2019crossover}; other symbols from Villois et al. \cite{villoisPRF2018}.\label{Fig:ApAm}}
\end{figure}

We evolve a Hopf link prepared by superimposing two rings of radius $ R=18\xi $, obtained by using a Newton--Raphson and biconjugate-gradient technique \cite{abid2003grossNewton} to ensure a minimal amount of compressible energy; details on the numerical scheme and on the generation of the initial condition are in the Supplemental Material.
%The computational box is periodic with sides of length $L=128\xi$; $256^3$ collocation points are used. 
%The initial Hopf link is prepared by superimposing two rings of radius $ R=18\xi $, each of them lying on a plane orthogonal to the other.
%We integrate numerically the GP model using a standard pseudo-spectral code evolved in time by a forth-order Runge--Kutta scheme. 
%The computational box is periodic with sides of length $L=128\xi$; $256^3$ collocation points are used. 
%The initial Hopf link is prepared by superimposing two rings of radius $ R=18\xi $, each of them lying on a plane orthogonal to the other.
%The order parameter of each ring is numerically obtained by using a Newton--Raphson and biconjugate-gradient technique \cite{abid2003grossNewton}, allowing to minimize the initial sound excitations in the system.
A set of 49 different realizations are obtained by changing the offsets $ (d_1, d_2) $ of one ring as sketched in Fig.\ref{fig1}(b), taking $ d_i \in [-9\xi, 9\xi] $ with unit step of $ 3\xi $.  
During the evolution one or more reconnection events occur.
It has been shown \cite{nazarenko2003analytical, villoisPRF2018, galantucci2019crossover, Enciso:2019aa} that about the reconnection event, the distance between the two filaments behaves as
\begin{equation}
\delta^{\pm}(t) = A^\pm (\Gamma |t-t_{\rm r}|)^{1/2} \, ,
\end{equation}
where $ A^\pm $ are dimensionless pre-factors and $ t_{\rm r} $ is the reconnection time; the superscripts $ - $ and $ + $ label the cases {\it before} and {\it after} the reconnection, respectively. 
In each Hopf link realization, we carefully track \cite{VilloisTrackingAlgo} all reconnecting events and measure $ A^\pm $.
The measured values of $ \delta^2(t) $ for all the 71 analyzed reconnections are shown in Fig.~\ref{Fig:ApAm}; the best-fit $ A^\pm $ are plotted in red dots in the inset of Fig.~\ref{Fig:ApAm}.
Remarkably, the reconnecting filaments always separate faster (or at an almost equal rate) than they approach, that is $ A^+ \ge A^- $.
The clear asymmetry recorded in the $ \delta^2 $ versus $ t-t_{\rm r} $ and in the distribution of the $ A^\pm $s is the fingerprint of the irreversible dynamics characterizing the vortex reconnection process.
For completeness, we also report in the inset of Fig.~\ref{Fig:ApAm}, using different symbols, the pre-factor measurements obtained in previous works \cite{villoisPRF2018, galantucci2019crossover}, which corroborate even further our results. Finally note that in a recent work \cite{yao_hussain_2020}, it has been reported that vortex reconnections in Navier-Stokes also display a clear $t^{1/2}$ scaling, a coefficient $A^-\sim 0.3-0.4$ and the same time asymmetry $A^+>A^-$.  
{Note that the Biot-Savart analytical calculations of reference \cite{Boue2013} and the LIA based ones of \cite{Rica:2019aa}, predict $A^-\sim0.47$ and $A^-=0.427$ respectively, which are in agreement with our GP measurements. }
In what follows, we quantitatively relate the asymmetry in the distribution of the pre-factors with the irreversible energy transfer between the vortex-type and density/phase excitation families occurring during a reconnection event. 
Previous numerical studies of the GP model have indeed reported the clear emission of a sound pulse during reconnection events \cite{PhysRevLett.86.1410, Zuccher&Caliari&Baggaley&BarenghiPof2012}.
A series of snapshots showing the sound pulse emitted during the decay of the Hopf link in one of our realizations is reported in \cite{Proment2020Matching}.

The simple linear theory neglecting the nonlinear term of the GP model \cite{nazarenko2003analytical, villoisPRF2018}, valid in the limit $\delta^\pm\to0$, provides an insight into the dynamics of reconnecting parameters as the the order parameter can be found analytically.
It predicts that the filaments reconnect tangent to a plane, in our reference frame the $ z=0 $, see Fig.~\ref{fig1}(a), and that the projections of the filaments onto it approach and separate following the branches of a hyperbola.
The {\it macroscopic (post) reconnection angle}, formed by the hyperbola asymptotes, results in
\begin{equation}
\phi^+=2\arccot(A_{\rm r}) \, , \quad \mbox{where $ A_{\rm r}=A^+/A^- $.}
\end{equation}
Moreover the projections of the filaments onto the orthogonal plane $ y=0 $ is a parabola (not shown in here, see \cite{Proment2020Matching} for more details).
Without any loss of generality, we set $ \Lambda/\zeta $ the concavity of such parabola, {and we refer to $ \Lambda $ as the {\it concavity parameter}, where $\zeta$ is an arbitrary length scale, which value is not important.} 

\begin{figure}
\includegraphics[width=0.5\textwidth]{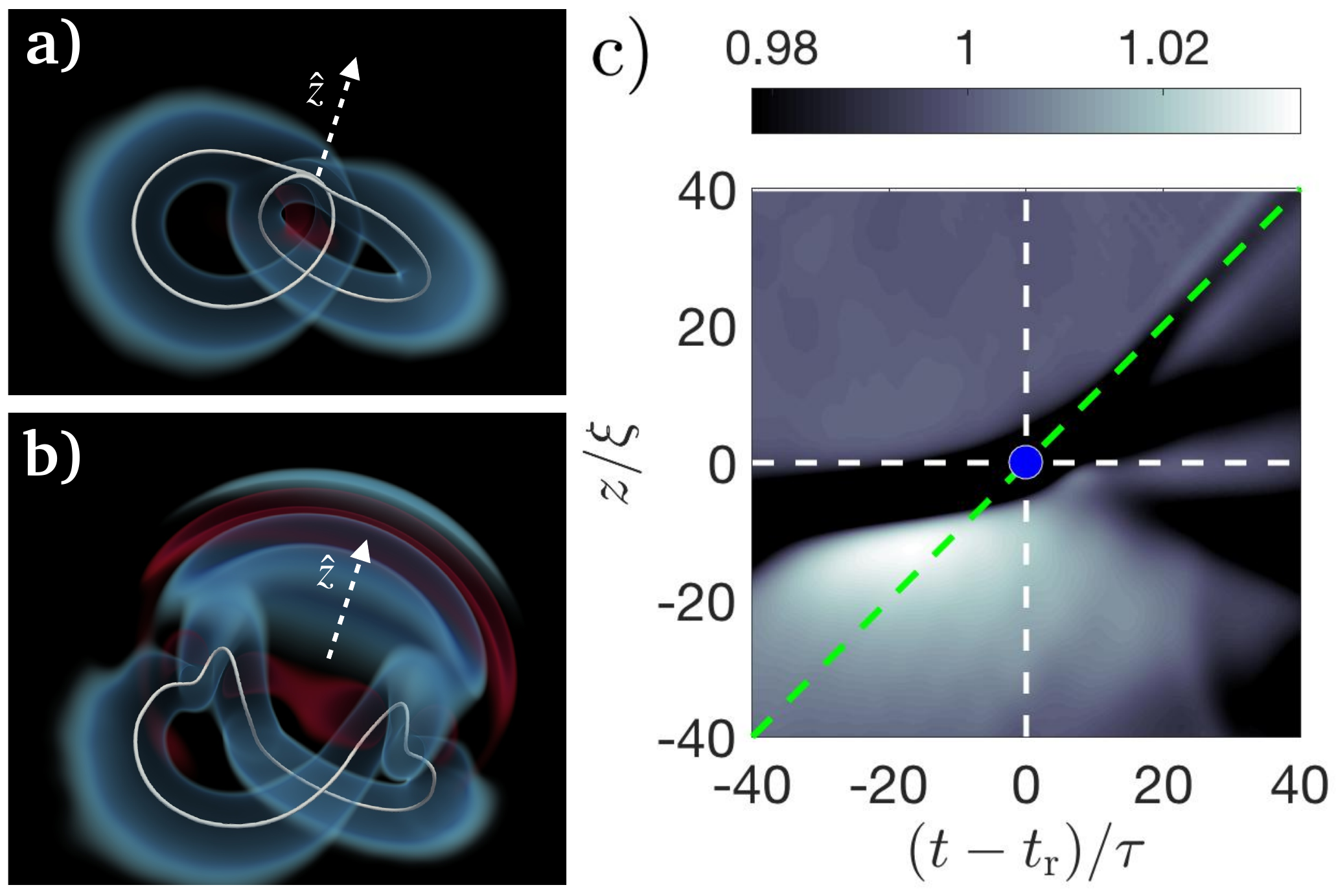} 
\caption{(Color online)
Three-dimensional rendering of the density field. White contours display the vortices and density fluctuations are rendered in blue-redish colors: (a) reconnection time and (b) at $t-t_{\rm r}\approx40\tau$ (b)
The positive direction of the $ z $-axis is also depicted with a white arrow.    
(c) Spatiotemporal plot of density along the $ z $-axis about the reconnection event denoted by the blue central point. The two dashed green lines are $z= c(t-t_{\rm r})$; here, the reconnection point $(0,0)$ is represented by the blue dot.
}
\label{Fig:Pulse}
\end{figure}

In all the reconnection events detected, we observe a neat sound pulse generated after the reconnection and propagating towards the positive $ z $-axis, as shown in Figs~\ref{Fig:Pulse}(a) and \ref{Fig:Pulse}(b).
Figure~\ref{Fig:Pulse}(c) shows the behavior of the superfluid density along the $ z $ direction versus times $ t-t_{\rm r} $.
A (depression) sound pulse is generated soon after the reconnection and propagates towards the positive $ z $-direction at a speed qualitatively close to the speed of sound in the bulk, refer to the green dashed line $z=c (t-t_{\rm r}) $, {with $c$ defined in Eq.~\eqref{Eq:GPhydro}}.
Note that the other low density regions, corresponding to the density depletions of the vortex cores, which move much slower.

To explain the generation and directionality of such pulse we devise a novel theoretical approach, detailed in \cite{Proment2020Matching}, and summarize in the following.
Let us denote by ${\bf R}^{\pm}_1(s,t)$ and ${\bf R}^{\pm}_2(s,t)$ the reconnecting filaments, with $ s $ being their spatial parametrization variable.
Far from the reconnection point (both before and after), the dynamics of the vortex filaments are mostly driven by the Biot-Savart (BS) model, which describes the motion of $ \delta$-supported vorticity in an incompressible inviscid flow \cite{SchwartzVFM}; note that this limit can be formally derived from GP \cite{BustamanteNazarenko}. 
In our realizations, BS is valid at distances $\delta^\pm(t) \gg \delta_{\rm lin} $, whereas for $\delta^\pm(t) \ll \delta_{\rm lin} $ the dynamics is determined by the linear approximation, given $ \delta_{\rm lin}$ is a crossover scale of order of the healing length. 
We assume both descriptions approximately valid when the filaments are at the distance $ \delta^\pm(t^\pm)\approx\delta_{\rm lin} $.
This hypothesis, validated by previous GP simulations \cite{villoisPRF2018,galantucci2019crossover}, allows us to perform an asymptotic matching.

We can therefore compute the difference, before and after the reconnection, of BS linear momentum $\Delta {\bf P}_{\rm fil}$ using the positions of the  filaments ${\bf R}^{\pm}_1(s, t^\pm)$ and ${\bf R}^{\pm}_2(s, t^\pm)$ coming from the linear approximation.
As shown in \cite{Proment2020Matching}, note that these depend only on the reconnection angle $ \phi^+ $ (or equivalently $A_{\rm r}$) and the concavity parameter $ \Lambda $.
Within BS, the linear momentum is given as the line integral ${\bf P}_{\rm fil}(t)=\frac{\rho_0}{2}\Gamma\oint{\bf R}(s, t)\times\mathrm{d}{\bf R}(s, t) $ \cite{Pismen:1999aa}.
As the total linear momentum of the superfluid is conserved in GP 
\footnote{This is formally true in a system which is invariant under spatial translations. We can extend this property to a finite system if we assume that the boundaries are sufficiently far from the reconnection point so that the conservation of the linear momentum is almost exact within a given volume enclosing the reconnection event.}, 
the linear momentum carried by the sound pulse created after the reconnection must compensate the loss of linear momentum accounted by $ \Delta {\bf P}_{\rm fil} $ and reads \cite{Proment2020Matching}
\begin{equation}
{\bf P}_{\rm pulse}=-\Delta {\bf P}_{\rm fil}\propto (0,0,2\csc{\phi^+}) \, ,
\label{Eq:DeltaP}
\end{equation}
independently on the $ \delta_{\rm lin} $ chosen.
This result is remarkable: the sound pulse linear momentum is (overall) non-zero only in the positive $ z $-direction, as observed in all our reconnections events, its amplitude is independent of $ \Lambda $ and minimal for {$\phi^+=\pi/2$}.

\begin{figure}
\includegraphics[width=0.5\textwidth]{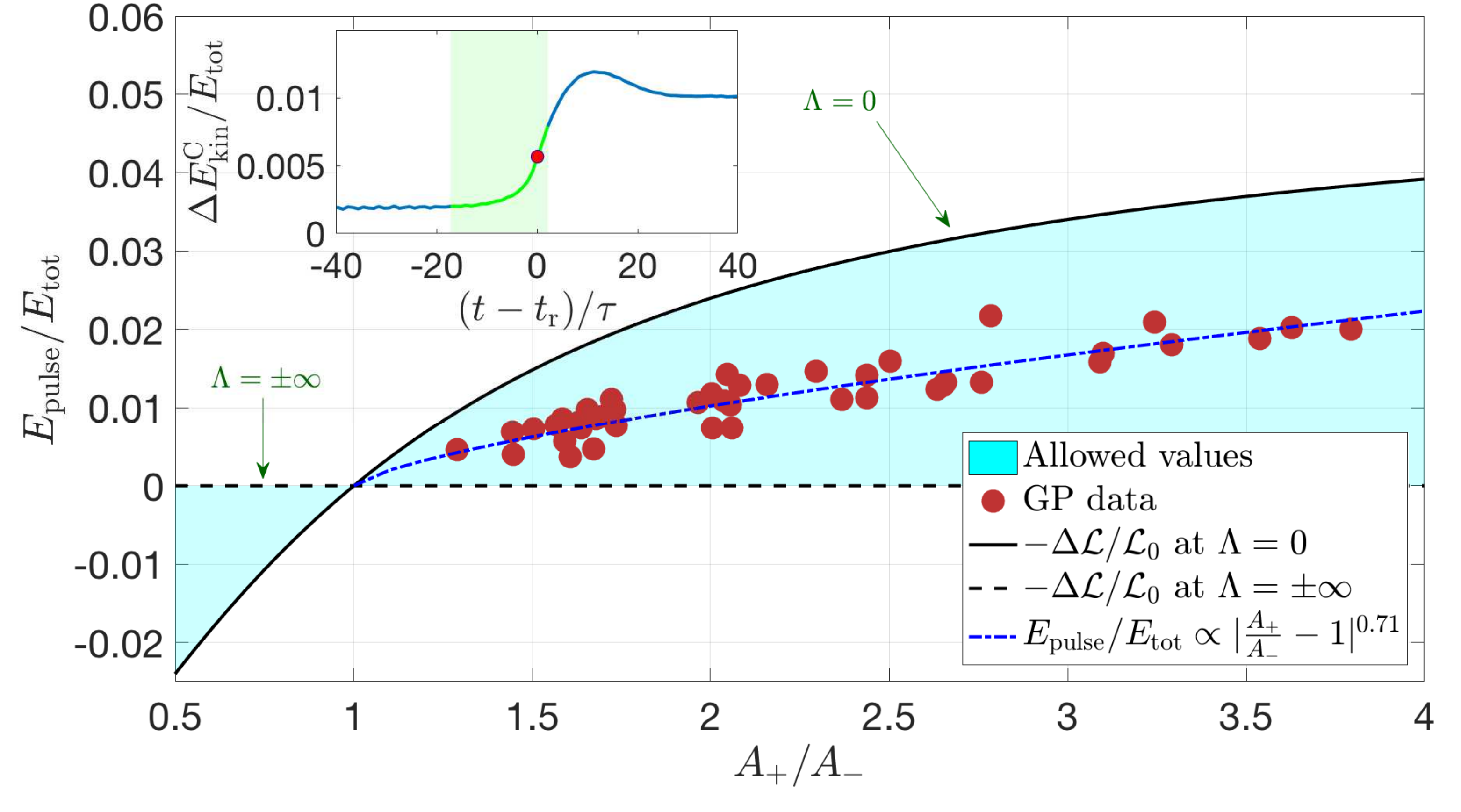} 
\caption{(Color online) 
Relative energy transferred to waves during the reconnection process. The cyan zone denotes the allowed values from the matching theory.
(inset) Relative increase of compressible kinetic energy (solid blue) about a reconnection event (denoted by the red dots) for a typical realization. The green area corresponds to the interval defined by $\delta^\pm(t)\le\delta_{\rm lin}=6\xi.$
\label{Fig:DeltaE}}
\end{figure}

The same matching theory can be applied to estimate the amount of energy transferred to the sound pulse.
Following the standard energy splitting protocol in GP \cite{nore1997decaying}, the superfluid kinetic energy is decomposed into a compressible component $ E_{\rm kin}^{\rm C} $, associated to sound excitations, and an incompressible component $ E_{\rm kin}^{\rm I} $, associated to vortex-type excitations. 
In all our realizations, we observe a sharp growth of $ E_{\rm kin}^{\rm C} $ during each reconnection event. 
An example of its evolution, normalized by the total (constant) energy $ E_{\rm tot} $,  is shown in the inset of Fig.~\ref{Fig:DeltaE}: here the red dot indicates the reconnection time, and the green region indicates the times when $ \delta^\pm(t) \le \delta_{\rm lin} = 6\xi $.  
The increase of $ E_{\rm kin}^{\rm C}$ during the reconnection event is related to the loss of incompressible kinetic energy $ E_{\rm kin}^{\rm I} $.
For all the reconnection events measured in our realizations, we compute the energy transferred to the sound pulse as $ E_{\rm pulse}=-\Delta E_{\rm kin}^{\rm I} $, where  $ \Delta E_{\rm kin}^{\rm I}=E_{\rm kin}^{\rm I}(t^+)-E_{\rm kin}^{\rm I}(t^-) $.
Figure~\ref{Fig:DeltaE} shows the measured $ E_{\rm pulse}/E_{\rm tot} $ data versus $ A_{\rm r} $: there is clear correlation between these two quantities, with a best-fit scaling of $ E_{\rm pulse} / E_{\rm tot} \propto (A_{\rm r}-1)^{0.71}$.

To the simplest approximation, called local induction approximation (LIA), the BS superfluid kinetic energy is proportional to the total length of the filaments. 
As the representations ${\bf R}_1(s, t)$ and ${\bf R}_2(s, t)$ have infinite lengths (as in the linear regime they do not close) we choose to account only for the length of finite sections of the filaments contained in a cylinder of radius $R\gg\delta_{\rm lin}$, centered at the reconnection point and parallel to the $ z $-axis.
Evaluating $ E_{\rm pulse}$ reduces thus to the computation of the difference $\Delta \mathcal{L}(A_{\rm r}, |\Lambda|/\zeta, \delta_{\rm lin}, R/\delta_{\rm lin}) $ of the length of these sections, see \cite{Proment2020Matching} for more details.
As the total GP energy is conserved, we have that
\begin{equation}
E_{\rm pulse}/E_{\rm tot}=-\Delta \mathcal{L}/\mathcal{L}_0 \, , 
\end{equation}
given $ \mathcal{L}_0 $ is the initial length of the Hopf link filament.
For any given choices of $ \delta_{\rm lin} $ and $ R $, all the admissible values of the theoretical estimation $ \Delta\mathcal{L} $, rendered in cyan color in Fig.~\ref{Fig:DeltaE}, are bounded between two lines obtained setting $ \Lambda=0 $ (dashed line) and $ |\Lambda| \to \infty $ (solid line).
The GP data are all distributed within these admissible values, thus confirming the accuracy of the matching theory.

Remarkably, the estimation of $ E_{\rm pulse} $ explains in a straightforward way the time asymmetry between the rates of approach and separation reported in Fig.~\ref{Fig:ApAm} and its inset.
Independently on the value of the concavity parameter $ \Lambda $, the energy of the sound pulse is only non-negative when $ A^+\ge A^- $, meaning that unless energy is externally provided to the reconnecting vortices, it is energetically impossible to have a reconnection event where $ A^+<A^- $, or equivalently, where $ \phi^+>\pi/2 $. stu

{\it Closing remarks}.
In this Letter we reported numerical evidence of the irreversible dynamics of vortex reconnections in a scalar quantum fluid, and explain its origin thanks to a matching theory developed concurrently in \cite{Proment2020Matching}. {This theory is based on very on general physical consideration and give bounds for the energy of the pulse emitted during a reconnection event. However, it cannot determine the exact value of the reconnecting angle and thus, the one of $A_+/A_-$.}
Our results can be extended to more complicated quantum fluids where non-local interactions and/or higher order nonlinearities are included, like BECs with dipolar interactions, cold Fermi gases, and superfluid liquid $ ^4 $He.

In quantum fluid experiments, the detailed study of vortex reconnections is still in its infancy.
In current BECs made of dilute gases, reconnecting vortices are created only in a non-reproducible way using fast temperature quenches \cite{PhysRevLett.115.170402}; however new protocols have been proposed to create vortices in a reproducible manner \cite{Xhani:2020aa}.
In such setups, once the reconnection plane is identified, it should be feasible to measure the rates of approach and separation and detecting directionality of the sound pulse, using for instance Bragg spectroscopy \cite{Steinhauer:2002aa}.
In superfluid liquid $ ^4 $He experiments, vortex reconnections have been detected so far only at relatively high temperature where the normal component is non-negligible \cite{BewleyReconnectionPNAS}. 
This latter may provide energy but also dissipate it through mutual friction, hence measuring experimentally the distribution of the rates of approach and separation at different temperatures would be particularly desirable.

Finally, let us come back to the concept of irreversibility.
In the realizations presented in this Letter, almost all of the superfluid kinetic energy is initially stored in the vortex-type excitations.
This is likely to cause the observed statistical asymmetry in the distribution of the rates of approach and separation to be maximized.
At finite temperatures or in a turbulent tangle, fluctuations can provide extra energy to reduce this asymmetry, perhaps allowing also for $ \phi^+>\pi/2 $, but the time-asymmetry should in principle remains as an inherent mechanism allowing the system to reach the equilibrium.
From a fluid dynamical point of view, let us to remark that vortex reconnections are allowed and regular, in classical fluids due to the presence of viscosity, while in quantum fluids thanks to a dispersive term.
Showing whether the resulting dynamics of these two different fluids are equivalent or not, in the limit where their respective regularization scale tends to zero, is an appealing open problem. 
Comparing the results presented in this Letter with a similar study in Navier--Stokes or a carefully regularized Biot-Savart model might provide some insights on the spontaneous stochasticity and the dissipative anomaly of turbulent flows, two concepts closely related to irreversibility.

\begin{acknowledgments}  
The authors acknowledge L.~Galantucci for providing some of the data displayed in the inset of Fig.~\ref{Fig:ApAm}.
G.K., D.P. and A.V. were supported by the cost-share Royal Society International Exchanges Scheme (IE150527) in conjunction with CNRS.
A.V. and D.P. were supported by the EPSRC First Grant scheme (EP/P023770/1).  
G.K. and D.P. acknowledge the Federation Doeblin for supporting D.P. during his sojourn in Nice. 
G.K. was also supported by the ANR JCJC GIANTE ANR-18-CE30-0020-0.1 and by the EU Horizon 2020 research and innovation programme under the grant agreement No 823937 in the framework of Marie Skodowska-Curie HALT project.
Computations were carried out at M\'esocentre SIGAMM hosted at the Observatoire de la C\^ote d'Azur and on the High Performance Computing Cluster supported by the Research and Specialist Computing Support service at the University of East Anglia.
Part of this work has been presented at the workshop ``Irreversibility and Turbulence''  hosted by Fondation Les Treilles in September 2017. 
G.K. and D.P. acknowledge Fondation Les Treilles and all participants of the workshop for the frightful scientific discussions and support.
\end{acknowledgments}

\bibliographystyle{aipnum4-1}
%\bibliography{../../ReconnectionsANDSound.bib}

\end{document}